\documentclass[12pt,final]{IEEEtran}
\usepackage{amsmath}
\usepackage{graphicx}
\usepackage{amsfonts}
\usepackage{amssymb}
\begin{document}

\title{How High Can The U-CAS Fly?}
\author{Shahar Gov and Shmuel Shtrikman, \emph{Fellow, IEEE}\thanks{The authors are
with the Department of Electronics, Weizmann Institute of Science, 76100
Rehovot, Israel (E-mail: fegov@weizmann.weizmann.ac.il). S. Shtrikman is also
with the Department of Physics, University of California, San Diego, La Jolla,
92093 CA, USA.}}
\date{}
\maketitle
\begin{abstract}
The U-CAS is a spinning magnetized top that is levitated in a static magnetic
field. The field is produced by a permanent magnet base, positioned below the
hovering top. In this paper we derive upper and lower bounds for $h_{m}$--the
maximum hovering height of this top. We show that the bounds are of the form
$al_{0}$ where $a$ is a dimensionless number ranging from about $1$ to $12$,
depending on the constraints on the shape of the base and on stability
considerations, and $l_{0}$ is a characteristic length, given by
\[
l_{0}\equiv\dfrac{\mu_{0}}{4\pi}\dfrac{\mu M_{0}}{mg}\text{.}%
\]
Here, $\mu_{0}$ is the permeability of the vacuum, $\mu$ is the magnetic
moment of the top, $M_{0}$ is the maximum magnetization of the base, $m$ is
the mass of the top, and $g$ is the free-fall acceleration. For modern
permanent magnets we find that $l_{0}\sim1$[meter], thus limiting $h$ to about
few meters.

\end{abstract}

\section{Introduction\label{sec1}}

\subsection{What is the U-CAS?\label{sec1.1}}

\bigskip The U-CAS is an ingenious device that hovers in mid-air while
spinning. It is marketed as a kit in Japan under the trade name U-CAS
\cite{ucas}, and in the U.S.A. and Europe under the trade name
Levitron$^{\text{TM}}$ \hspace{0pt} \cite{levitron,harrigan,patent}. The whole
kit consists of three main parts: A magnetized top which weighs about $18$ gr,
a thin (lifting) plastic plate and a magnetized square base plate (base). To
operate the top one should set it spinning on the plastic plate that covers
the base. The plastic plate is then raised slowly with the top until a point
is reached in which the top leaves the plate and spins in mid-air above the
base for about 2 min. The hovering height of the top is approximately $3$ cm
above the surface of the base whose dimensions are about 10 cm $\times$ 10 cm
$\times$ 2 cm. The kit comes with extra brass and plastic fine tuning weights,
as the apparatus is very sensitive to the weight of the top. It also comes
with two wedges to balance the base horizontally.

\subsection{The adiabatic approximation.\label{sec1.2}}

The physical principle underlying the operation of the U-CAS relies on the
so-called `adiabatic approximation' \cite{bergeman,berry,coils}: As the top is
launched, its magnetic moment points \emph{antiparallel} to the magnetization
of the base in order to supply the repulsive magnetic force which will act
against the gravitational pull. As the top hovers, it experiences lateral
oscillations which are slow ( $\Omega_{\text{lateral}}\simeq1$ Hz) compared to
its precession ($\Omega_{\text{precession}}\sim5$ Hz). The latter itself, is
small compared to the top's spin ($\Omega_{\text{spin}}\sim25$ Hz). Since
$\Omega_{\text{spin}}\gg\Omega_{\text{precession}}$ the top is considered
`fast' and acts like a classical spin. Furthermore, as $\Omega
_{\text{precession}}\gg\Omega_{\text{lateral}}$ this spin may be considered as
experiencing a \emph{slowly} rotating magnetic field. Under these
circumstances the spin precesses around the \emph{local} direction of the
magnetic field $\mathbf{B}$ (adiabatic approximation) and, on the average, its
magnetic moment $\mathbf{\mu}$ points \emph{antiparallel} to the local
magnetic field lines. In view of this discussion, the magnetic interaction
energy which is normally given by $-\mathbf{\mu}\cdot\mathbf{B}$ is now given
approximately by $\mu\left|  \mathbf{B}\right|  $. Thus, the overall effective
energy `seen' by the top is
\begin{equation}
E_{\text{eff}}\left(  \mathbf{r}\right)  \simeq mgz+\mu\left|  \mathbf{B}%
\left(  \mathbf{r}\right)  \right|  , \label{energy}%
\end{equation}
where $m$ is the mass of the top, $g$ is the free-fall acceleration and $z$ is
the height of the top above the base. By virtue of the adiabatic
approximation, two of the three rotational degrees of freedom are coupled to
the transverse translational degrees of freedom, and as a result the rotation
of the axis of the top is already incorporated in Eq.(\ref{energy}). Thus,
under the adiabatic approximation, the top may be considered as a
\emph{point-like} particle whose only degrees of freedom are translational.
The important point of this discussion is the following: The energy expression
written above \emph{possesses a minimum} for certain values of $\mu/mg$. Thus,
when the mass is properly tuned, the apparatus acts as a trap, and stable
hovering becomes possible. A detailed description of this device, extending
beyond the adiabatic approximation, may be found elsewhere
\ \cite{berry,simon,dynamic}. For the purpose of this paper, the adiabatic
approximation will suffice.

\subsection{The limitations on the hovering height.\label{sec1.3}}

In this paper we focus on the question:
\[
\text{How high can the hovering height of the U-CAS be?}%
\]
First, we define what do we mean by `height': We assume that the top hovers
above some horizontal plane. The height of the top is measured with respect to
that plane. The `rules of the game' are, that we can put permanent magnet
\emph{below} the plane, but never \emph{above }this plane. Whatever is below
this plane, will be called `the base'. The method we use to answer the
question above will be by getting upper and lower \emph{bounds} for this
height, denoted by $h_{m}$. We begin by pointing out the factors that limit
the hovering height of the U-CAS.

First, we set aside the question of stability and assume that the top is
guided by a vertical axis, allowing it to move only axially (up and down). In
this case, the easiest way to increase $h$--the hovering height, is by using a
more powerful magnet for the base. This technique however, cannot be applied
indefinitely since there is no way to increase the magnetization of a given
substance without limit. We are therefore forced to limit the strength of
$\mathbf{M(r)}$--the magnetization density of the base, to some maximum value,
say $M_{0}$, and design the shape and magnetization density of the base so as
to maximize $h$. One more factor that limits the levitation height is the
amount of magnetic substance, or \emph{volume}, in our disposal. Clearly, the
more material we use, the larger is the levitation height that can be
achieved. But still, even if the volume is infinite the levitation height is
bounded, since $\left|  \mathbf{M(r)}\right|  $ is bounded.

When stability is taken into account, things get more complicated. Stability
sets another limitation on the design of the base, in \emph{addition} to the
limitations that were discussed above: A brief look at Eq.(\ref{energy}) tells
us that for the top to hover \emph{stably} over the plate, the effective
energy $E_{\text{eff}}$ should possess a \emph{minimum}. This means, in
particular, that
\begin{equation}
\dfrac{\partial^{2}E_{\text{eff}}\left(  \mathbf{r}\right)  }{\partial
\mathbf{r}^{2}}\sim\dfrac{\partial^{2}\left|  \mathbf{B}\left(  \mathbf{r}%
\right)  \right|  }{\partial x^{2}},\dfrac{\partial^{2}\left|  \mathbf{B}%
\left(  \mathbf{r}\right)  \right|  }{\partial y^{2}},\dfrac{\partial
^{2}\left|  \mathbf{B}\left(  \mathbf{r}\right)  \right|  }{\partial z^{2}%
}>0\text{,} \label{stab}%
\end{equation}
where the derivatives are evaluated at the equilibrium position of the top.
Since these are \emph{homogenous} inequalities, it is clear that the region in
space where a minimum \emph{may} occur, does not depend on the \emph{strength}
of the magnetic field. As a consequence, the hovering height of the U-CAS\ is
\emph{not} determined by the strength of the magnetic field but on its
\emph{geometry, }or alternatively, by the \emph{shape} of the base. As an
example, it has been shown \cite{tolerance} that for a vertically magnetized
base in the shape of a disk of radius $R$, the range of heights $h$ for which
stable hovering is possible, is very narrow and is around $h\sim R/2$. This
result agrees roughly with the parameters of the U-CAS, for which $R\sim10$ cm
and $h\sim4$ cm, when $h$ is measured from the center-of-mass of the base. The
\emph{strength} of the field then comes into play by tuning it in order to
achieve equilibrium for a given mass of the top. Equilibrium prevails when the
total force on the top vanishes, i.e. when
\begin{equation}
{\mathbf{F}{=-\nabla}}E_{\text{eff}}=-mg{\mathbf{\hat{z}}}-\mu{\mathbf{\nabla
}}\left|  {\mathbf{B(r)}}\right|  =0\text{.} \label{force}%
\end{equation}

Both in the guided case and in the stable case, we see that in order to
increase the hovering height of the U-CAS, we should design a better base. We
do expect however, that the hovering height cannot be increased indefinitely.
In connection with the redesigning of the base, we have recently shown
\cite{ring}, both theoretically and experimentally, that the use of a
vertically magnetized \emph{ring} of radius $R$ as a base, increases the
hovering height by more than three times to about $1.7R$!. This increase in
height did not came without cost, as the tolerance on the mass of the top
became more stringent, being $\Delta m/m\sim0.6\%$ for the ring vs. $\Delta
m/m\sim1.2\%$ in the case of the disk \cite{tolerance}.

\subsection{The structure of this paper.\label{sec1.4}}

All the calculations that we do are outlined in Sec.(\ref{sec2}). We will now
describe what we calculate and what is the motivation behind it.

We start in Sec.(\ref{sec2.1}), by deriving a close form expression for the
first and second derivatives of the magnetic field, in terms of the
magnetization density of the base. The first derivative is the magnetic force
on the top, which is used throughout the paper, while the second derivative is
exploited when we discuss stability.

In Sec.(\ref{sec2.2}) we consider the problem of maximizing the hovering
height in the case where the \emph{volume} of the base may be \emph{infinite},
and may be \emph{arbitrarily} magnetized under the constraint that $\left|
\mathbf{M(r)}\right|  =M_{0}$. We still \emph{do not} require that the top be
stable against lateral translations, and assume that it is guided along a
vertical axis. We show that in this case
\[
h_{\max}\simeq12l_{0},
\]
where $l_{0}$ is a characteristic height for the problems that are discussed
in this paper. It is defined as
\[
l_{0}\equiv\dfrac{\mu_{0}}{4\pi}\dfrac{\mu M_{0}}{mg},
\]
where $\mu_{0}$ is the magnetic permeability of the vacuum, $\mu$ is the
magnetic moment of the top, $m$ is the mass of the top, and $g$ is the
free-fall acceleration.

In Sec.(\ref{sec2.3}) we consider the same problem as before but for a base
which is \emph{uniformly} magnetized along the vertical direction, namely
${\mathbf{M(r)}}=M_{0}{\mathbf{\hat{z}}}$, where $\mathbf{\hat{z}}$ is a unit
vector in the vertical direction. In this case we find that
\[
h_{\max}\simeq3.5l_{0}.
\]

In order to arrive to better bounds on $h_{m}$, our next step is to limit the
amount of material available for the base. Thus, in Sec.(\ref{sec2.4}) we
maximize the hovering height in the case where the base may be
\emph{arbitrarily} magnetized under the constraint that $\left|
\mathbf{M(r)}\right|  =M_{0}$, and that its \emph{volume} $V_{0}$ is given.
Here, the result is given in the form of a plot, showing the dependence
$h_{\max}$ on $V_{0}$. In the limit $V_{0}\rightarrow\infty$, we recover the
result of Sec.(\ref{sec2.2}). Our derivation is also \emph{constructive} in
the sense that it shows how to \emph{construct} the optimal base for a given
$V_{0}$. In this paper however, we do not discuss this matter thoroughly due
to space shortage.

For completeness, we study in Sec.(\ref{sec2.5}) the same problem as in
Sec.(\ref{sec2.4}), but for a \emph{uniformly} magnetized base. Here, again we
find the dependence of $h_{\max}$ on $V_{0}$ and recover the result of
Sec.(\ref{sec2.3}).

Sec.(\ref{sec2.6}) is the first part where stability is added into play: We
begin by explaining how to \emph{test} for stability of the top under the
adiabatic approximation, and study the possible levitation heights of a
\emph{uniformly} magnetized base whose shape is \emph{cylindrical} of volume
$V_{0}$. As the stability condition is given in the form of inequality
relations (see Eq.(\ref{stab})), it is not suffice to determine uniquely \ the
levitation height. We therefore choose to study two particular stable points:
an \emph{isotropic} stable point and the \emph{highest} possible stable point.
The meaning of these points would become clear later. In this case we also
give a plot, showing the dependence of $h_{\max}$ on $V_{0}$. This result
should also be considered as a lower bound for $h_{m}$, since the cylindrical
base is a special case of all the base configurations that are possible.

In section(\ref{sec2.7}) we take one step further, and generalize the result
of section(\ref{sec2.6}) to the case of a base which may be \emph{arbitrarily}
magnetized, and look for the \emph{lowest} possible stable point. In this case
however, we assume that $V_{0}=\infty$, to ease the solution of the problem.
We also show that in this case, where $h_{\max}\simeq8.4l_{0}$ and
$V_{0}=\infty$, the \emph{lowest} possible stable point and the \emph{highest}
one coincide, so this result represent not a \emph{bound }for $h_{m}$, but is
actually $h_{m}$ itself, for the case $V_{0}=\infty$.

In Sec.(\ref{sec3}) we discuss interesting aspects of our results. In
particular we estimate the value of $l_{0}$ for modern permanent magnet
materials and calculate the values of the bounds for $h_{m}$. We comment on
the implications of our results and discuss other related questions.

\section{Mathematical Formulation\label{sec2}}

\subsection{The upward magnetic force and its derivative.\label{sec2.1}}

\begin{figure}[t]
\begin{center}
\includegraphics[
natheight=7.338800in,
natwidth=6.859700in,
height=3.5907in,
width=3.3589in
]{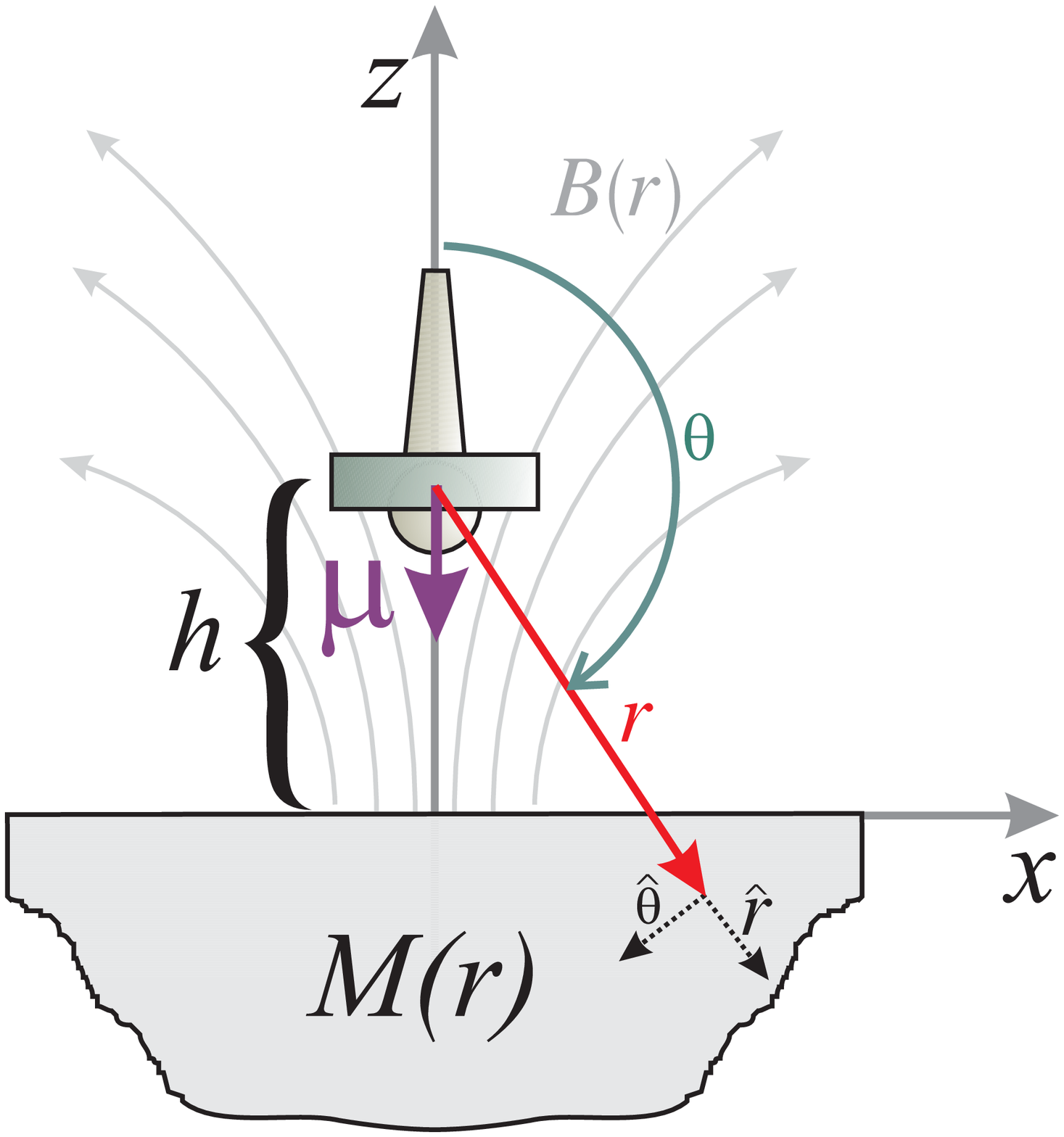}
\end{center}
\caption{The U-CAS is modeled as a point-like particle of mass $m$ and
magnetic moment $\mu$ pointing downward, hovering at a height $h$ above a
non-uniformly magnetized base with magnetization density $\mathbf{M(r)}$
producing a field $\mathbf{B(r)}$. }%
\label{fig1}%
\end{figure}

A simplified model of the U-CAS is shown in Fig.(\ref{fig1}). It consists of a
point-like particle of mass $m$ and magnetic moment $\mu$ (pointing downward),
hovering at a height $h$ above the $z=0$ plane. For the moment, we consider
the top as if it was \emph{guided} along the $z$-axis so that only its
vertical motion is allowed. Other degrees of freedom are considered `frozen'.
The region $z<0$ may be partially or wholly filled with a magnetized substance
(``base''), whose magnetization density is denoted by $\mathbf{M(r)}$,
producing a magnetic field throughout space which we denote by $\mathbf{B(r)}%
$. In what follows, we assume that the base (and hence the magnetic field) has
a cylindrical symmetry around the $z$-axis. Thus, along this axis the magnetic
field possesses only a $z$-component. We further assume that this component is
\emph{positive}, namely that the magnetic field along the $z$-axis is pointing upward.

Using Eqs.(\ref{energy}),(\ref{force}) we find that, when the top is in
equilibrium, the total vertical component of the force on the top vanishes,
\[
{\mathbf{F}}_{z}=-mg{\mathbf{\hat{z}}}-\mu\dfrac{\partial\left|  {\mathbf{B}%
}\right|  }{\partial z}{\mathbf{\hat{z}}}=0,
\]
where the term $-mg{\mathbf{\hat{z}}}$ is the gravitational force pulling the
top towards the base, and the term $-\mu{\mathbf{\hat{z}}}\partial\left|
{\mathbf{B}}\right|  /\partial z$ is the magnetic force which pushes the top
upward. Note that $\partial\left|  {\mathbf{B}}\right|  /\partial z$ should be
\emph{negative}.

Since the top is allowed to move only along the $z$-axis, we may write
$\left|  \mathbf{B}\right|  =B_{z}$ (in what follows we deserve the notation
$B_{z}$ to denote the $z$-component of the magnetic field along the $z$-axis,
namely $B_{z}\equiv{\mathbf{\hat{z}\cdot B}}\left(  x=0,y=0,z\right)  $). Now,
the magnetic force on the top simplifies to
\[
{\mathbf{F}}_{m}=-\mu\frac{\partial B_{z}}{\partial z}{\mathbf{\hat{z}},}%
\]
and hence, when the top is in equilibrium at $z=h$,
\begin{equation}
-\mu\left.  \frac{\partial B_{z}}{\partial z}\right|  _{z=h}=mg. \label{eq1.1}%
\end{equation}

Reciprocity allows us to express $\partial B_{z}/\partial z$ in terms of
$\mathbf{M(r)}$ as%

\begin{equation}
\frac{\partial B_{z}}{\partial z}=\int{\mathbf{M(r}}^{\prime})\cdot
\frac{\partial{\mathbf{B}}_{d}({\mathbf{r}}^{\prime};z)}{\partial z}%
d^{3}r^{\prime}. \label{eq1.2}%
\end{equation}
Here, ${\mathbf{B}}_{d}({\mathbf{r}}^{\prime};z)$ is the field at the point
$\mathbf{r}^{\prime}$ produced by a unit magnitude dipole pointing
\emph{upward}, located at a height $z$ along the $z$-axis. Taking $r$ and
$\theta$ as depicted in Fig.(\ref{fig1}), we can write ${\mathbf{B}}_{d}$ as
\begin{equation}
{\mathbf{B}}_{d}=\frac{\mu_{0}}{4\pi}\frac{(3\cos\theta{\mathbf{\hat{r}%
-\hat{z}}})}{r^{3}}, \label{eq1.3}%
\end{equation}
where $\mathbf{\hat{r}}$ is a polar unit vector, also defined in
Fig.(\ref{fig1}), and $\mu_{0}$ is the magnetic permeability of the vacuum. It
is worth to note that $\partial{\mathbf{B}}_{d}/\partial z$ is nothing but the
field produced at $\mathbf{r}^{\prime}$ by a \emph{quadruple} located at a
height $z$ along the $z$-axis. This quadruple is made out of a pair of
identical dipoles: One is located at $z$ and pointing downward and the other
is at an infinitesimally higher position $z+dz$ and pointing upward, with
their magnetic moment being $1/dz$.

When Eq.(\ref{eq1.3}) is substituted into Eq.(\ref{eq1.2}) we find that, at
$z=h$, the field derivative is given by
\begin{equation}
\left.  \frac{\partial B_{z}}{\partial z}\right|  _{z=h}=\frac{\mu_{0}}{4\pi}%
%TCIMACRO{\dint \limits_{\pi/2}^{\pi}}%
%BeginExpansion
{\displaystyle\int\limits_{\pi/2}^{\pi}}
%EndExpansion
\sin\theta d\theta%
%TCIMACRO{\dint \limits_{-h/\cos(\theta)}^{\infty}}%
%BeginExpansion
{\displaystyle\int\limits_{-h/\cos(\theta)}^{\infty}}
%EndExpansion
2\pi r^{2}dr\frac{1}{r^{4}}{\mathbf{M}}(r,\theta)\cdot\overline{{\mathbf{n}}%
}(\theta),\label{eq1.4}%
\end{equation}
where
\begin{equation}
\overline{\mathbf{n}}(\theta)\equiv6\sin\theta\cos\theta{\mathbf{\hat{\theta}%
}}+3(3\cos^{2}\theta-1){\mathbf{\hat{r}},}\label{n}%
\end{equation}
and where ${\mathbf{\hat{\theta}}}$ is a polar unit vector orthogonal to
$\mathbf{\hat{r}}$, as is shown in Fig.(\ref{fig1}).

We would also need the \emph{second} derivative of the field, when we discuss
stability. It is given by
\begin{equation}
\left.  \frac{\partial^{2}B_{z}}{\partial z^{2}}\right|  _{z=h}=\frac{\mu_{0}%
}{4\pi}%
%TCIMACRO{\dint \limits_{\pi/2}^{\pi}}%
%BeginExpansion
{\displaystyle\int\limits_{\pi/2}^{\pi}}
%EndExpansion
\sin\theta d\theta%
%TCIMACRO{\dint \limits_{-h/\cos(\theta)}^{\infty}}%
%BeginExpansion
{\displaystyle\int\limits_{-h/\cos(\theta)}^{\infty}}
%EndExpansion
2\pi r^{2}dr\frac{1}{r^{5}}{\mathbf{M}}(r,\theta)\cdot\overline{\mathbf{s}%
}(\theta), \label{eq1.41}%
\end{equation}
where
\begin{equation}
\overline{\mathbf{s}}(\theta)\equiv9\sin\theta\left(  5\cos^{2}\theta
-1\right)  {\mathbf{\hat{\theta}}}+12\cos\theta(5\cos^{2}\theta-3)\mathbf{\hat
{r}.} \label{s}%
\end{equation}
Note that in both Eqs.(\ref{eq1.4}) and (\ref{eq1.41}), ${\mathbf{M}}%
(r,\theta)$ is defined with respect to a polar coordinate system whose origin
is at $z=h$ and \emph{not }at $z=0$. Also note, that in going from
Eq.(\ref{eq1.4}) to Eq.(\ref{eq1.41}), one should take into account the
dependence of the unit vectors ${\mathbf{\hat{\theta}}}$ and $\mathbf{\hat{r}%
}$ on $z$.

\subsection{Arbitrarily magnetized, infinite base.\label{sec2.2}}

We consider an infinitesimal volume element at some point $(r,\theta)$ within
the base. It is essentially a magnetic dipole whose magnitude is $M_{0}$, and
whose direction we will now find by the requirement that the magnetic force on
the top is maximized. We have already shown that the \emph{force }on the top,
contributed by one elemental dipole, is proportional to the magnetic field
${\mathbf{B}}_{d}$ that would be produced at $(r,\theta)$ by a quadruple
located at $z=h$. To make this force maximal we assign to each point
$(r,\theta)$ in $z<0$, a magnetization density $\mathbf{M(r)}$ whose magnitude
is $M_{0}$ and whose direction is \emph{antiparallel} to the field line of
${\mathbf{B}}_{d}$ at that point. The magnetization density so defined will
\emph{maximize} the force on the top, and is therefore the requested answer.
Mathematically, this procedure amounts to replacing $\mathbf{M(r)}$ in
Eq.(\ref{eq1.4}) by ${\mathbf{M(r)}}=-M_{0}{\mathbf{\hat{n}}}(\theta)$, where
${\mathbf{\hat{n}}}(\theta)$ is a unit vector \emph{parallel} to
$\overline{\mathbf{n}}(\theta)$, the latter being defined in Eq.(\ref{n}). The
result is
\begin{align*}
\left.  \dfrac{\partial B_{z}}{\partial z}\right|  _{z=h}  &  =-\dfrac{\mu
_{0}}{4\pi}6\pi M_{0}%
%TCIMACRO{\dint \limits_{\pi/2}^{\pi}}%
%BeginExpansion
{\displaystyle\int\limits_{\pi/2}^{\pi}}
%EndExpansion
d\theta\sin\theta\sqrt{4\cos^{4}\theta+\sin^{4}\theta}\\
&  \times%
%TCIMACRO{\dint \limits_{-h/\cos(\theta)}^{\infty}}%
%BeginExpansion
{\displaystyle\int\limits_{-h/\cos(\theta)}^{\infty}}
%EndExpansion
\dfrac{1}{r^{2}}dr,
\end{align*}
for which the $r$ integration is trivial and the $\theta$ integration may be
brought to a simpler form by the transformation $x=\sin^{2}(\theta)$. This
gives \cite{rem1}
\[
\left.  \dfrac{\partial B_{z}}{\partial z}\right|  _{z=h}=-\dfrac{\mu_{0}%
}{4\pi}\dfrac{3\pi M_{0}}{h}%
%TCIMACRO{\dint \limits_{0}^{1}}%
%BeginExpansion
{\displaystyle\int\limits_{0}^{1}}
%EndExpansion
\sqrt{4(1-x)^{2}+x^{2}}dx\cong-\dfrac{12M_{0}}{h}\dfrac{\mu_{0}}{4\pi},
\]
which together with Eq.(\ref{eq1.1}) shows that
\begin{equation}
h_{\max}\cong12l_{0}\text{,} \label{eq1.6}%
\end{equation}
where
\[
l_{0}\equiv\dfrac{\mu_{0}}{4\pi}\dfrac{M_{0}\mu}{mg},
\]
is the characteristic length in our problem which will reappear in the next
sections. The value of $l_{0}$ may be interpreted as the \emph{distance
}between two colinear dipoles, one of them is of strength $\mu$ while the
other is of strength $M_{0}l_{0}^{3}$, for which the mutual force between them
is $mg$.

Eq.(\ref{eq1.6}) shows that even though the magnetization direction is allowed
to vary everywhere inside the base, the levitation height $h$ is bounded.
Eq.(\ref{eq1.6}) presents the maximum height that can be accomplished with a
given substance provided that the top is not allowed to move laterally.
Clearly, it also serve as an \emph{upper }bound for $h_{m}$, as stability was
not considered yet. Moreover, note that we have calculated the maximum
magnetic force, which is proportional to $\partial B_{z}/\partial z$. The
magnetic field $\left.  B_{z}\right|  _{z=h}$ on the other hand, becomes
\emph{infinite} at $z=h,$ as can be seen by the following simple argument:
Since each elemental dipole within the plate contributes a field that goes as
$1/r^{3}$ and since the volume of integration goes as $r^{2}$, the integrand
goes as $1/r$, for which the integral diverges as $r\rightarrow\infty$. Later
we show that the divergence of $\left.  B_{z}\right|  _{z=h}$ implies that the
top \emph{cannot} be stable when placed in such a point. In any case, it would
be impossible to \emph{spin} the top.

\subsection{Uniformly magnetized, infinite base.\label{sec2.3}}

Uniformly magnetized plates are clearly easier to construct. In this section
we find out what upper bound does this restriction sets on the highest
levitation point. In another sense, the result of this section may also be
considered as a \emph{lower} bound on the height of levitation (for a top
which is guided!), provided one considers all possible base configurations.

We again consider an infinitesimal volume element within the base. It is now
oriented along the $z$-direction and interacting with the top's magnetic
dipole. If this element is exactly below the top (i.e. it is located somewhere
on the negative\textbf{\ }$z$-axis), it exerts a repelling $z$-directed force
on the top, pushing it away. If the element is not exactly below the top then
the nature of this force (i.e. weather it is repulsive or attractive) is
determined by the angle formed between the direction of the two dipoles (in
this case both point in the $z$-direction), and the direction defined by the
line joining these dipoles. We call this angle $\theta$, and measure it with
respect to the positive $z$-direction, as is shown in Fig.\ref{fig1}.

\begin{figure}[th]
\begin{center}
\includegraphics[
natheight=11.008200in,
natwidth=11.008200in,
height=3.6236in,
width=3.4022in
]{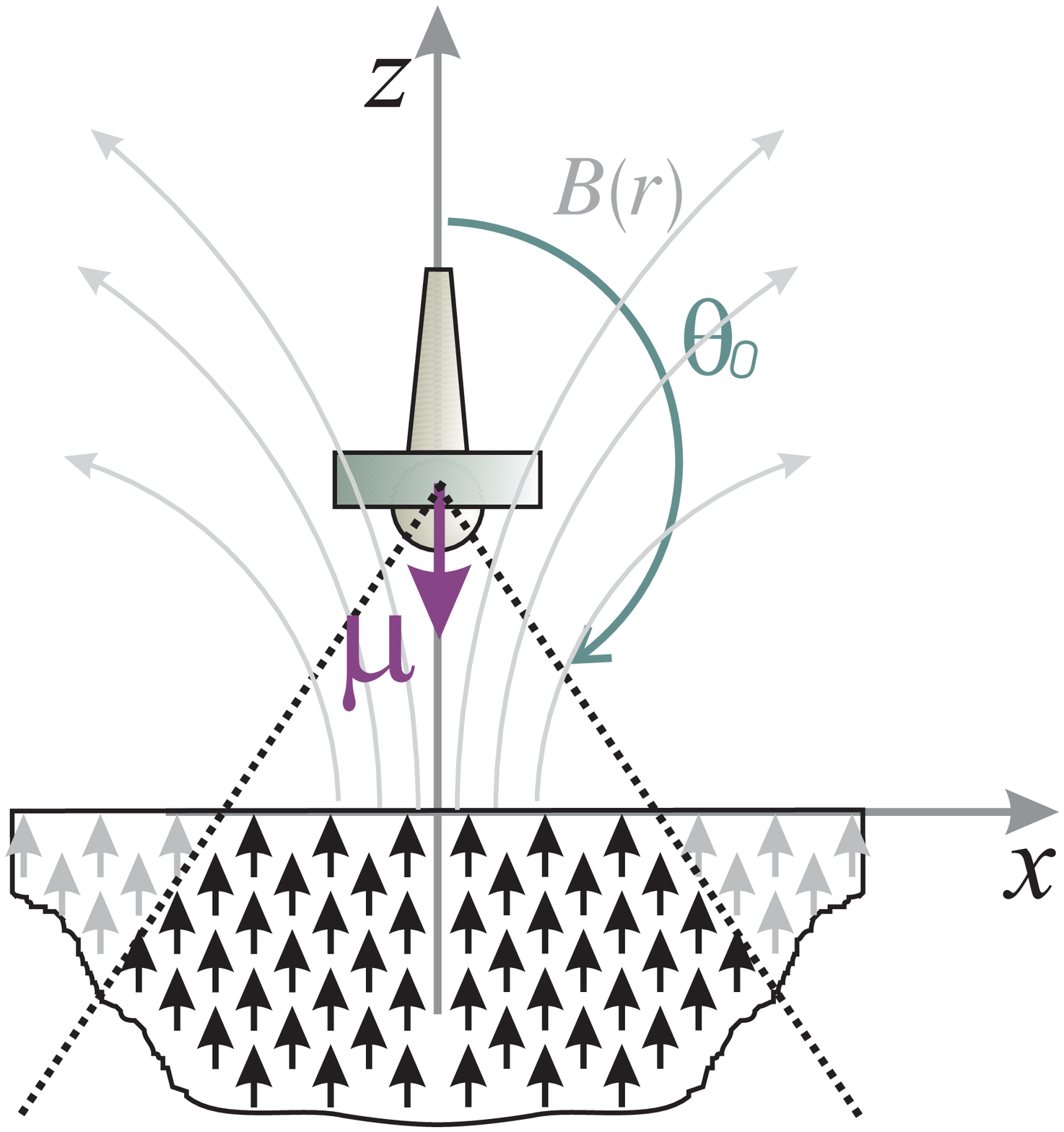}
\end{center}
\caption{For a uniformly magnetized base, all the dipoles that lie within the
angular sector $\theta_{0}<\theta<(2\pi-\theta_{0})$ (black dipoles), where
$\theta_{0}\simeq141^{0}$, \emph{repel }the top. The dipoles that are outside
this sector (grey dipoles) \emph{attract} the top. Material should be put
\emph{only} in the repulsive sector. }%
\label{fig2}%
\end{figure}

There exists a critical angle $\theta_{0}$ (see Fig.(\ref{fig2})), for which
the $z$-directed force vanishes such that for $\pi/2<\theta<\theta_{0}$ the
force becomes attractive. It is therefore useless to put any material within
the region $\pi/2<\theta<\theta_{0}$ and $(2\pi-\theta_{0})<\theta<3\pi/2$
since this would \emph{reduce} the magnetic force and decrease the levitation
height.

To find $\theta_{0}$ we eliminate the $z$-component of Eq.(\ref{n}), and find
the value of $\theta$ for which it vanishes. We pick the value of $\theta$
that lies between $\pi/2$ and $\pi$. This gives
\[
\theta_{0}=\pi-\arccos\sqrt{3/5}\cong141^{o}.
\]

We now evaluate the upward force exerted on the top, by summing the
contributions of \emph{all} the elements in the base located within
$\theta_{0}<\theta<(2\pi-\theta_{0})$. The (uniform) magnetization density
within this region is taken to be ${\mathbf{M(r)}}=M_{0}{\mathbf{\hat{z}}}$.
Using Eq.(\ref{eq1.4}), we find that
\begin{align*}
\left.  \dfrac{\partial B_{z}}{\partial z}\right|  _{z=h}  &  =\frac{\mu_{0}%
}{4\pi}%
%TCIMACRO{\dint \limits_{\theta_{0}}^{\pi}}%
%BeginExpansion
{\displaystyle\int\limits_{\theta_{0}}^{\pi}}
%EndExpansion
\sin\theta d\theta%
%TCIMACRO{\dint \limits_{-h/\cos(\theta)}^{\infty}}%
%BeginExpansion
{\displaystyle\int\limits_{-h/\cos(\theta)}^{\infty}}
%EndExpansion
2\pi r^{2}dr\frac{1}{r^{4}}(M_{0}{\mathbf{\hat{z}}})\\
&  \cdot\lbrack6\sin\theta\cos\theta{\mathbf{\hat{\theta}}}+3(3\cos^{2}%
\theta-1)\mathbf{\hat{r}}]\text{.}%
\end{align*}
The integration over $r$ is again trivial, and we are left with an integration
over $\theta$. The latter is brought to a simpler form by changing the
$\theta$ variable into $x=\cos\theta$, hence
\begin{align*}
\left.  \dfrac{\partial B_{z}}{\partial z}\right|  _{z=h}  &  =-\frac{\mu_{0}%
}{4\pi}\frac{6\pi M_{0}}{h}%
%TCIMACRO{\dint \limits_{-1}^{-\sqrt{3/5}}}%
%BeginExpansion
{\displaystyle\int\limits_{-1}^{-\sqrt{3/5}}}
%EndExpansion
(5x^{4}-3x^{2})dx\\
&  =-\frac{\mu_{0}M_{0}}{h}\left(  \dfrac{3}{5}\right)  ^{5/2}.
\end{align*}
This result, together with the equilibrium condition Eq.(\ref{eq1.1}),
suggests that in this case the maximal levitation height possible is
\begin{equation}
h_{\max}=4\pi\left(  \dfrac{3}{5}\right)  ^{5/2}\left[  \dfrac{M_{0}\mu}%
{mg}\dfrac{\mu_{0}}{4\pi}\right]  \equiv4\pi\left(  \dfrac{3}{5}\right)
^{5/2}l_{0}\cong3.5l_{0}\text{.} \label{eq1.5}%
\end{equation}

Note, however, that in order to realize the levitation heights found in this
section and in the previous one we would need an unlimited supply of magnetic
material. In the following sections we find how good can we do when the
\emph{volume} of the base is constrained.

\subsection{Arbitrarily magnetized, finite base.\label{sec2.4}}

Under a given volume $V_{0}$, we find the optimum \emph{shape} and
\emph{magnetization} of the base, that will maximize the levitation height. We
use the Lagrange's multipliers method to treat this variational problem
\cite{lag}.

\begin{figure}[th]
\begin{center}
\includegraphics[
natheight=7.338800in,
natwidth=6.859700in,
height=3.6244in,
width=3.3892in
]{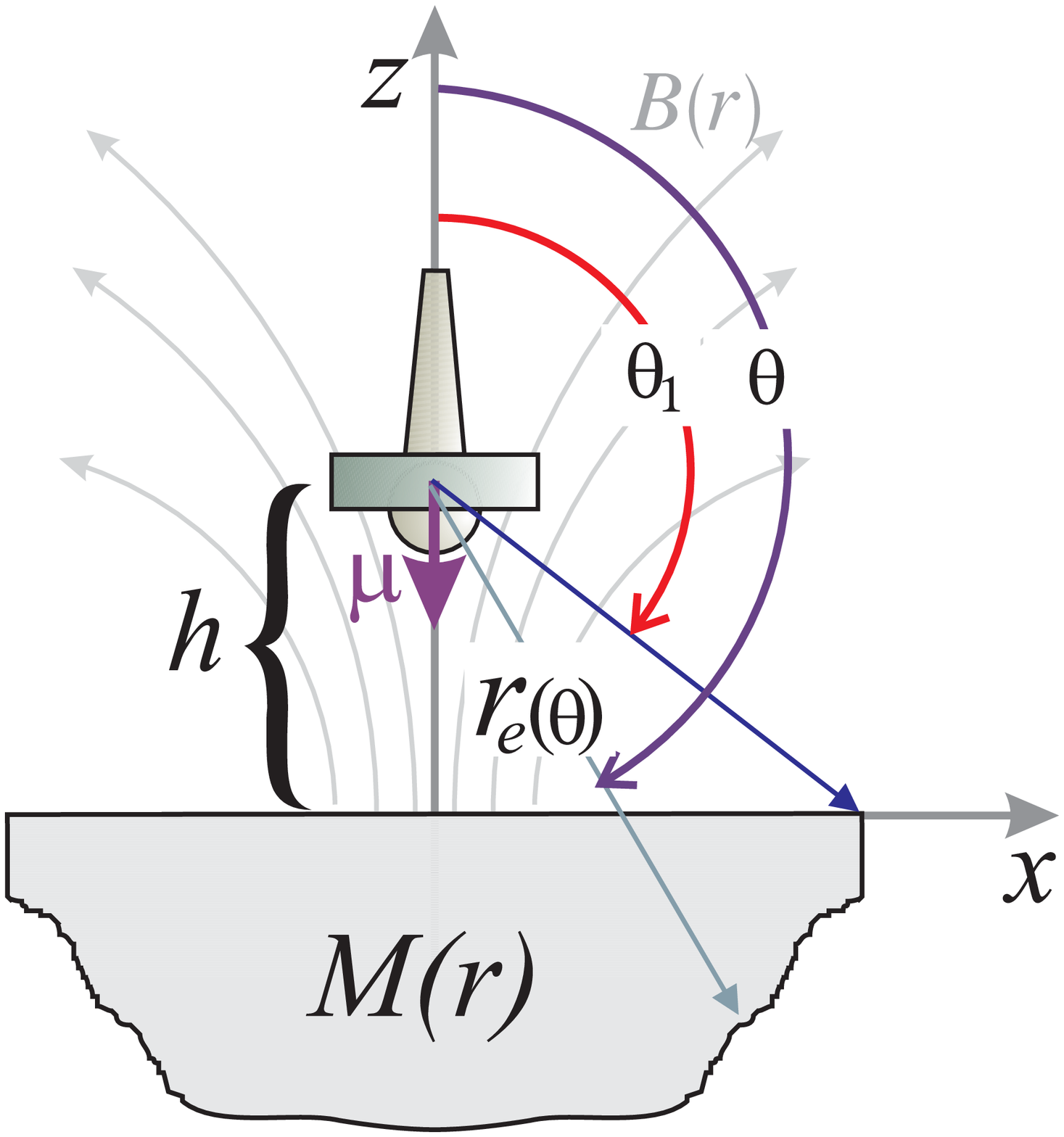}
\end{center}
\caption{A \emph{finite} base is parametrized by $r_{e}(\theta)$-the radial
distance to the boundary for a given $\theta$. Integration in $r$ is carried
out from $r=-h/\cos\theta$ to $r_{e}\left(  \theta\right)  $. The polar angle
of the upper-right corner of the base is $\theta_{1}$.}%
\label{fig3}%
\end{figure}

First, we parametrize the \emph{shape} of the plate: Let $r_{e}(\theta)$ be
the upper integration radius, as is shown in Fig.(\ref{fig3}). The lower
integration radius is $-h/\cos\theta$. Utilizing the cylindrical symmetry of
the problem, the volume may be written as%

\begin{equation}%
\begin{array}
[c]{c}%
V=2\pi%
%TCIMACRO{\dint \limits_{\theta_{1}}^{\pi}}%
%BeginExpansion
{\displaystyle\int\limits_{\theta_{1}}^{\pi}}
%EndExpansion
\sin\theta d\theta%
%TCIMACRO{\dint \limits_{-h/\cos\theta}^{r_{e}(\theta)}}%
%BeginExpansion
{\displaystyle\int\limits_{-h/\cos\theta}^{r_{e}(\theta)}}
%EndExpansion
r^{2}dr\\
=\dfrac{2\pi}{3}%
%TCIMACRO{\dint \limits_{\theta_{1}}^{\pi}}%
%BeginExpansion
{\displaystyle\int\limits_{\theta_{1}}^{\pi}}
%EndExpansion
d\theta\sin\theta\left(  r_{e}^{3}(\theta)+\dfrac{h^{3}}{\cos^{3}\theta
}\right)  ,
\end{array}
\label{eq1.7}%
\end{equation}
where the angle $\theta_{1}$ is determined by the condition that
\begin{equation}
r_{e}(\theta_{1})=-h/\cos\theta_{1},\label{eq1.8}%
\end{equation}
and is the value of the angle to the upper-right corner of the base, as is
shown in Fig.(\ref{fig3}).

The upward force Eq.(\ref{eq1.4}), which in this case is given by setting
\[
{\mathbf{M(r)}}=M_{0}{\mathbf{\hat{n}}(\theta),}%
\]
is obtained from
\begin{align}
\dfrac{1}{M_{0}}\dfrac{4\pi}{\mu_{0}}\left.  \dfrac{\partial B_{z}}{\partial
z}\right|  _{z=h}  &  =-6\pi%
%TCIMACRO{\dint \limits_{\theta_{1}}^{\pi}}%
%BeginExpansion
{\displaystyle\int\limits_{\theta_{1}}^{\pi}}
%EndExpansion
d\theta\sin\theta\sqrt{4\cos^{4}\theta+\sin^{4}\theta}\label{eq1.9}\\
&  \times\left(  -\frac{1}{r_{e}(\theta)}-\dfrac{\cos\theta}{h}\right)
.\nonumber
\end{align}

According to the Lagrange's multipliers method \cite{lag}, the target function
that is to be maximized is
\begin{equation}
T\equiv\frac{1}{M_{0}}\dfrac{4\pi}{\mu_{0}}\left.  \dfrac{\partial B_{z}%
}{\partial z}\right|  _{z=h}-\frac{1}{\lambda^{4}}\left(  V_{0}-V\right)  ,
\label{eq1.10}%
\end{equation}
where $1/\lambda^{4}$ is the Lagrange's multiplier and $V_{0}$ is the given
volume of the base. The target function is a functional of $r_{e}(\theta)$. We
require that a variation of $T$ with respect to $r_{e}(\theta)$ would vanish,
thus
\begin{equation}
\frac{\delta T}{\delta r_{e}(\theta)}=\frac{1}{M_{0}}\dfrac{4\pi}{\mu_{0}%
}\frac{\delta}{\delta r_{e}(\theta)}\left[  \left.  \dfrac{\partial B_{z}%
}{\partial z}\right|  _{z=h}\right]  +\frac{1}{\lambda^{4}}\frac{\delta
V}{\delta r_{e}(\theta)}=0. \label{eq1.11}%
\end{equation}
Using Eqs.(\ref{eq1.8}),(\ref{eq1.9}) we find that
\begin{equation}
\frac{\delta T}{\delta r_{e}(\theta)}=2\pi%
%TCIMACRO{\dint \limits_{\theta_{1}}^{\pi}}%
%BeginExpansion
{\displaystyle\int\limits_{\theta_{1}}^{\pi}}
%EndExpansion
d\theta\sin\theta\left\{  -\frac{3\sqrt{4\cos^{4}\theta+\sin^{4}\theta}}%
{r_{e}^{2}(\theta)}+\frac{r_{e}^{2}(\theta)}{\lambda^{4}}\right\}  \text{.}
\label{eq1.12}%
\end{equation}
Therefore, the required parametrization $r_{e}(\theta)$ is obtained by
equating the integrand in Eq.(\ref{eq1.12}) to zero, giving
\begin{equation}
\frac{r_{e}(\theta)}{\lambda}=\left[  3\sqrt{4\cos^{4}\theta+\sin^{4}\theta
}\right]  ^{1/4}. \label{eq1.13}%
\end{equation}
Eq.(\ref{eq1.13}) defines the \emph{shape} of the optimal base. We see that it
is universal in the sense that as $V_{0}$ is changed, the optimal new plate's
shape is only a scaled version of the original one. Recall however, that
$\theta_{1}$ is \emph{different}.

The value of $\theta_{1}$, determined by Eq.(\ref{eq1.8}), may be combined
with Eq.(\ref{eq1.13}) to read
\begin{equation}
\frac{h}{\lambda}=-\cos\theta_{1}\left[  3\sqrt{4\cos^{4}\theta_{1}+\sin
^{4}\theta_{1}}\right]  ^{1/4}=f^{-1}\left(  \theta_{1}\right)  ,
\label{eq1.14}%
\end{equation}
which implicitly expresses $\theta_{1}$ in terms of $h/\lambda$. We denote
this function by $f(x)$, and write
\[
\theta_{1}=f(h/\lambda)\text{.}%
\]
In addition, we use Eq.(\ref{eq1.13}) to arrive to an explicit expression for
the volume $V_{0}$, given in Eq.(\ref{eq1.7}):
\[
\dfrac{V_{0}}{\lambda^{3}}=G(h/\lambda),
\]
where%
\[
G\left(  x\right)  \equiv\dfrac{2\pi}{3}%
%TCIMACRO{\dint \limits_{\theta_{1}=f(x)}^{\pi}}%
%BeginExpansion
{\displaystyle\int\limits_{\theta_{1}=f(x)}^{\pi}}
%EndExpansion
d\theta\sin\theta\left(
\begin{array}
[c]{c}%
\left[  3\sqrt{4\cos^{4}\theta+\sin^{4}\theta}\right]  ^{3/4}\\
+\dfrac{x^{3}}{\cos^{3}\theta}%
\end{array}
\right)  \text{.}%
\]
Similarly, applying Eq.(\ref{eq1.13}) to Eq.(\ref{eq1.9}) yields
\begin{equation}
\left.  \dfrac{\partial B_{z}}{\partial z}\right|  _{z=h}\lambda
=-M_{0}S(h/\lambda) \label{eq1.16}%
\end{equation}
with the definition%
\begin{align*}
S(x)  &  \equiv6\pi%
%TCIMACRO{\dint \limits_{\theta_{1}=f(x)}^{\pi}}%
%BeginExpansion
{\displaystyle\int\limits_{\theta_{1}=f(x)}^{\pi}}
%EndExpansion
d\theta\sin\theta\sqrt{4\cos^{4}\theta+\sin^{4}\theta}\\
&  \times\left[  \dfrac{-1}{\left(  3\sqrt{4\cos^{4}\theta+\sin^{4}\theta
}\right)  ^{1/4}}-\dfrac{\cos\theta}{x}\right]  \text{.}%
\end{align*}
Using Eq.(\ref{eq1.16}) with the equilibrium condition Eq.(\ref{eq1.1}), we
find that
\begin{equation}
\frac{\lambda}{l_{0}}=S(h/\lambda), \label{eq1.17}%
\end{equation}
and with few more steps we arrive to
\begin{equation}%
\begin{array}
[c]{c}%
\dfrac{h}{l_{0}}=\dfrac{h}{\lambda}S\left(  \dfrac{h}{\lambda}\right)  ,\\
\dfrac{V_{0}}{l_{0}^{3}}=G\left(  \dfrac{h}{\lambda}\right)  S^{3}\left(
\dfrac{h}{\lambda}\right)  .
\end{array}
\label{eq1.18}%
\end{equation}

We see that $h/l_{0}$ depends on $V_{0}/l_{0}^{3}$ through an intermediate
variable $h/\lambda$. It can be solved numerically by ``running'' over a wide
range of $h/\lambda$ and evaluating $h/l_{0}$ and $V_{0}/l_{0}^{3}$ for each
of its values. The result is given by the dash-dotted line in Fig.(\ref{plots}).

Note that at the limit $V_{0}\rightarrow\infty$ we find that $h=12l_{0}$, in
agreement with the result of section(\ref{sec2.2}). The asymptotic behavior of
$h$ as $V_{0}\rightarrow0$ is quite interesting also: This limit may be
evaluated from the above equations but it is much simpler (and more
instructive) to use the following argument: At the low volume limit, the base
may be considered as a dipole centered at the origin. Such an assumption is
valid only if $(h/l_{0})^{3}\gg V_{0}/l_{0}^{3}$. In this case it is easy to
see that the magnetic force, acting on the top, is just
\[
\left.  \dfrac{\partial B_{z}}{\partial z}\right|  _{z=h}=-\dfrac{\mu_{0}%
}{4\pi}\dfrac{6M_{0}V_{0}}{h^{4}},
\]
and hence
\[
-\mu\left.  \dfrac{\partial B_{z}}{\partial z}\right|  _{z=h}=\dfrac{\mu_{0}%
}{4\pi}\mu\dfrac{6M_{0}V_{0}}{h^{4}}=mg.
\]
Using the definition of $l_{0}$ we may rewrite the last result as
\[
\dfrac{h}{l_{0}}=\left[  6\dfrac{V_{0}}{l_{0}^{3}}\right]  ^{1/4}.
\]
We thus conclude that
\begin{equation}
\lim\limits_{V_{0}/l_{0}^{3}\rightarrow0}\left(  \dfrac{h}{l_{0}}\right)
\rightarrow\left[  6\dfrac{V_{0}}{l_{0}^{3}}\right]  ^{1/4}. \label{eq1.19}%
\end{equation}
Note that since
\[
\lim_{V_{0}/l_{0}^{3}\rightarrow0}\dfrac{V_{0}/l_{0}^{3}}{(h/l_{0})^{3}}%
\sim\left(  \frac{V_{0}}{l_{0}^{3}}\right)  ^{1/4}=0,
\]
we see that our assumption is confirmed. The asymptotic line of
Eq.(\ref{eq1.19}) is also plotted in Fig.(\ref{plots}) by the dash-dot-dotted
line. It is conspicuous that it is indeed an asymptotic.

\subsection{Uniformly magnetized, finite base.\label{sec2.5}}

The solution for this case is essentially similar to the solution presented in
the previous section. The only difference is in the form of the magnetization
that is used. Here we take ${\mathbf{M(r)}}=M_{0}{\mathbf{\hat{z}}}$ instead
of ${\mathbf{M(r)}}=M_{0}{\mathbf{\hat{n}}}(\theta)$. The relation between
$h/l_{0}$ and $V_{0}/l_{0}^{3}$ is again given by Eqs.(\ref{eq1.18}) with the
following new definitions:
\[%
\begin{array}
[c]{c}%
f^{-1}(\theta)\equiv\cos\theta\left[  3\cos\theta\left(  5\sin^{2}%
\theta-2\right)  \right]  ^{1/4},\\
G(x)\equiv\dfrac{2\pi}{3}%
%TCIMACRO{\dint \limits_{f(x)}^{\pi}}%
%BeginExpansion
{\displaystyle\int\limits_{f(x)}^{\pi}}
%EndExpansion
d\theta\sin\theta\left(
\begin{array}
[c]{c}%
\left[  3\cos\theta\left(  5\sin^{2}\theta-2\right)  \right]  ^{3/4}\\
+\dfrac{x^{3}}{\cos^{3}\theta}%
\end{array}
\right)  ,\\
S(x)\equiv6\pi%
%TCIMACRO{\dint \limits_{f(x)}^{\pi}}%
%BeginExpansion
{\displaystyle\int\limits_{f(x)}^{\pi}}
%EndExpansion
d\theta\sin\theta\cos\theta\left(  5\sin^{2}\theta-2\right) \\
\times\left[  \dfrac{-1}{\left[  3\cos\theta\left(  5\sin^{2}\theta-2\right)
\right]  ^{1/4}}-\dfrac{\cos\theta}{x}\right]  .
\end{array}
\]

The result of this calculation is given by the dotted line in Fig.(\ref{plots}%
). Note that in this case $h\rightarrow3.5l_{0}$ as $V_{0}\rightarrow\infty$,
which is again in agreement with the result of section(\ref{sec2.3}). The
asymptotic behavior of $h$ as $V_{0}\rightarrow0$ is identical to the result
that was found in the previous section.

\subsection{Uniformly magnetized, finite, cylindrical base and
stability.\label{sec2.6}}

\begin{figure}[th]
\begin{center}
\includegraphics[
natheight=7.433100in,
natwidth=7.587000in,
height=3.5907in,
width=3.6633in
]{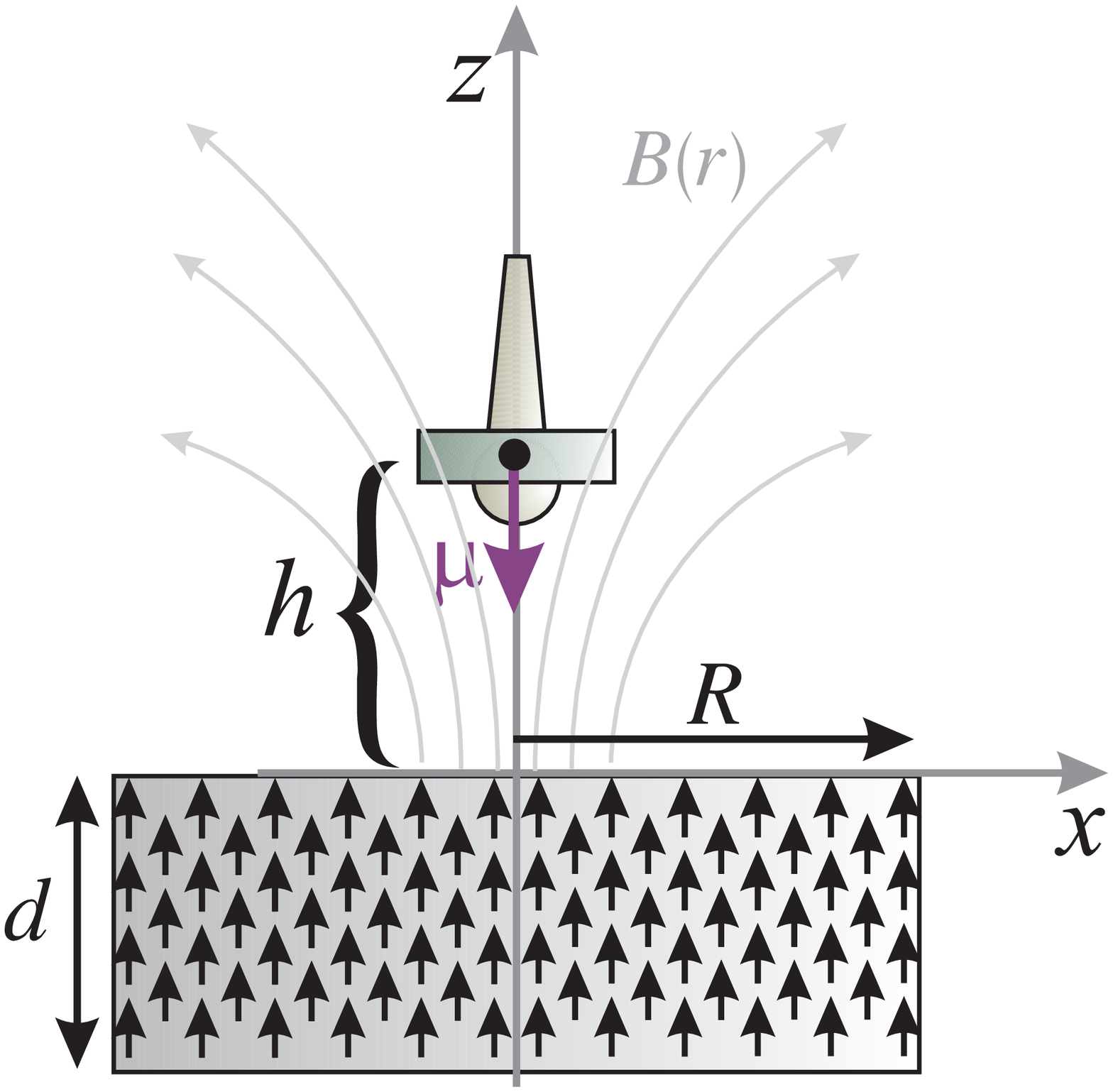}
\end{center}
\caption{The top hovers \emph{stably} over a cylindrical base of radius $R$,
thickness $d$, and vertical magnetization $M_{0}$. }%
\label{fig4}%
\end{figure}

In this case we take the base to be a uniformly magnetized cylinder with a
magnetization ${\mathbf{M(r)}}=M_{0}{\mathbf{\hat{z}}}$. The radius of the
cylinder is $R$ and its thickness is $d$, with its upper base at the $z=0$
plane, as is shown in Fig.(\ref{fig4}).

The magnetic field outside the cylinder is essentially the field outside a
solenoid with similar dimensions. Thus, the $z$-component of the magnetic
field along the $z$-axis is given by\cite{purcell}:
\begin{equation}
B_{z}(z)=\dfrac{\mu_{0}}{4\pi}M_{0}W\left(  \dfrac{z}{R};\dfrac{d}{R}\right)
, \label{eq1.27}%
\end{equation}
where%
\[
W\left(  x,y\right)  \equiv2\pi\left[  \dfrac{x}{\sqrt{x^{2}+1}}-\dfrac
{x+y}{\sqrt{(x+y)^{2}+1}}\right]  \text{.}%
\]
We also define the function $F(x,y)$, to be used later, as%
\[
F(x,y)\equiv-\dfrac{\partial W\left(  x,y\right)  }{\partial x}\text{.}%
\]

We now formulate the conditions for the \emph{stability} of the spinning top
against both axial and lateral translations: Stability in the $z$-direction is
determined by the sign of the `spring-constant' in that direction -- $k_{z}$.
Under the adiabatic approximation, the latter is given by the \emph{curvature}
of the effective energy along that direction. Hence,
\begin{equation}
k_{z}\equiv\left.  \dfrac{\partial^{2}E_{\text{eff}}}{\partial z^{2}}\right|
_{\substack{z=h\\\rho=0}}=\mu\left.  \frac{\partial^{2}\left|  \mathbf{B}%
\right|  }{\partial z^{2}}\right|  _{\substack{z=h\\\rho=0}}=\mu\left.
\frac{\partial^{2}B_{z}}{\partial z^{2}}\right|  _{z=h}, \label{eq1.30}%
\end{equation}
where $\rho$ is the radial distance, in cylindrical coordinate system, from
the $z$-axis.

Similarly, stability in the lateral direction $\hat{\rho}$, is governed by the
sign of
\begin{align}
k_{\rho}  &  \equiv\left.  \dfrac{\partial^{2}E_{\text{eff}}}{\partial\rho
^{2}}\right|  _{\substack{z=h\\\rho=0}}=\mu\left.  \frac{\partial^{2}\left|
\mathbf{B}\right|  }{\partial\rho^{2}}\right|  _{\substack{z=h\\\rho
=0}}\label{eq1.31}\\
&  =\mu\left[  \frac{1}{4\left.  B_{z}\right|  _{z=h}}\left(  \left.
\frac{\partial B_{z}}{\partial z}\right|  _{z=h}\right)  ^{2}-\frac{1}%
{2}\left.  \frac{\partial^{2}B_{z}}{\partial z^{2}}\right|  _{z=h}\right]
,\nonumber
\end{align}
where in the last equality use has been made of the cylindrical symmetry of
the magnetic field and the fact that $B_{z}$ and all of its Cartesian
derivatives are Harmonic functions \cite{jack}.

For the spinning top to be stable against translations, both $k_{z}$ and
$k_{\rho}$ should be \emph{positive}. Comparing Eq.(\ref{eq1.31}) to
Eq.(\ref{eq1.30}), we see that when $B_{z}\rightarrow\infty$, then $k_{\rho}$
$\rightarrow-$ $k_{z}/2$, and therefore one of the pair $(k_{z}$,$k_{\rho})$
\emph{must} be negative. Thus, when the magnetic field diverges at a point, a
top placed at that point cannot be stable. This proves that the highest
hovering height, found in Section (\ref{sec2.2}), is not under stable conditions.

The restriction that both $k_{z}$ and $k_{\rho}$ should be positive defines a
stable \emph{region} along the $z$ axis. As an example it can be shown
\cite{dynamic}, that for a base in the shape of a thin disk of radius $R_{d}$,
the value of $k_{z}$ is positive whenever $z>R_{d}/2$, whereas $k_{\rho}$ is
positive for $z<\sqrt{2/5}R_{d}$. The region of stability in this case is
therefore $R_{d}/2<z<\sqrt{2/5}R_{d}$. Within that region there exists a point
$z_{i}$ for which $k_{z}=k_{\rho}$. In the case of the disk it is $z_{i}%
=\sqrt{2/7}R_{d}$. We call this point the \emph{isotropically stable }point
because the restoring (stabilizing) force which acts on a top, which is tuned
to hover at $h=z_{i}$ is isotropic, depending only on the deviation from the
equilibrium position and not on its direction.

For a general cylinder we now consider two distinct situations: The first is
the one in which the stable point is \emph{isotropic}, i.e. a hovering height
$h$ for which $k_{\rho}=k_{z}$. The second is the case where $h$ is at the
verge of stability in the lateral direction. This is also the \emph{highest}
stable point, characterized by $k_{\rho}=0$. Using Eqs.(\ref{eq1.30}%
),(\ref{eq1.31}) we write each of the two distinct conditions as
\begin{equation}
\frac{1}{\left.  B_{z}\right|  _{z=h}}\left(  \left.  \frac{\partial B_{z}%
}{\partial z}\right|  _{z=h}\right)  ^{2}=a\left.  \frac{\partial^{2}B_{z}%
}{\partial z^{2}}\right|  _{z=h}, \label{eq1.32}%
\end{equation}
where $a=6$ for the first situation and $a=2$ for the second.

Substituting Eq.(\ref{eq1.27}) into Eq.(\ref{eq1.32}) defines a functional
relationship between $h/R$ and $d/R$ denoted by the function $G()$,%
\[
\frac{h}{R}\equiv G\left(  d/R\right)  .
\]
Differentiating Eq.(\ref{eq1.27}) with respect to $z$, setting $z=h$, and
using the equilibrium condition Eq.(\ref{eq1.1}), gives
\[
\frac{R}{l_{0}}=F\left(  h/R;d/R\right)  =N\left(  d/R\right)  ,
\]
where%
\[
N(y)\equiv F(G(y),y)\text{.}%
\]
Combining it with an expression for the volume, gives
\begin{equation}%
\begin{array}
[c]{c}%
\dfrac{V_{0}}{l_{0}^{3}}=\dfrac{\pi R^{2}d}{l_{0}^{3}}=\pi\left(  \dfrac{d}%
{R}\right)  \left(  \dfrac{R}{l_{0}}\right)  ^{3}=\pi\dfrac{d}{R}N^{3}\left(
d/R\right)  ,\\
\dfrac{h}{l_{0}}=\dfrac{h}{R}\dfrac{R}{l_{0}}=G(d/R)N\left(  d/R\right)  ,
\end{array}
\label{eq1.33}%
\end{equation}
which expresses the volume $V_{0}$ and the levitation height $h$ in terms of a
common variable $d/R$. Running over $d/R$, and evaluating the volume and
height according to Eqs.(\ref{eq1.33}), furnishes the required relation
between $h/l_{0}$ and $V_{0}/l_{0}^{3}$. This plot is shown in
Fig.(\ref{plots}) where the solid line corresponds to $a=2$ (the isotropic
case) and the dotted line corresponds to $a=6$ (the highest stable point).

Note that both of these plots are \emph{not} monotonically increasing. They
possess a maximum of $h$ at some optimal volume $V_{opt}$. For $a=6$ we find
that $h_{\max}\simeq1.3l_{0}$ and $V_{opt}\simeq310l_{0}^{3}$, whereas for
$a=2$ these are $h_{\max}\simeq0.88l_{0}$ and $V_{opt}\simeq90l_{0}^{3}$. This
indicates that using too a much material \emph{worsen} the largest height that
can be achieved, which is reminiscent of our conclusion of Sec.(\ref{sec2.3}).
If the given volume is larger than $V_{opt}$ however, we can always use only
an amount of volume equal to $V_{opt}$ and discard the rest of the material.
Hence, in principal at least, one \emph{can} realize the largest possible
height, which is why the plots of $h$ vs. $V_{0}$ had been artificially
corrected by assigning the maximum value of $h$ for the values of $V_{0}$ that
are larger than $V_{opt}$.

\subsection{Infinite,arbitrarily magnetized base and stability.\label{sec2.7}}

In this last section we consider the case of an infinite base, which may be
arbitrarily magnetized, such that the height of levitation is maximized, yet
the top is stable against both axial and lateral translations. In order to
solve this problem, one needs to maximize the magnetic force $\partial
B_{z}/\partial z$ under the constraints that $k_{z}$ and $k_{\rho}$, defined
in Eqs.(\ref{eq1.30}) and (\ref{eq1.31}) respectively, are \emph{both}
positive. The last requirement however, results in a non-linear inequality,
which cannot be solved analytically. The method we take here is to use the
constraint $k_{z}\propto\partial^{2}B_{z}/\partial z^{2}=0$ instead, which
marks the lower end of the stability region along the $z$-axis.

\begin{figure}[th]
\begin{center}
\includegraphics[
height=3.5907in,
width=3.3581in
]{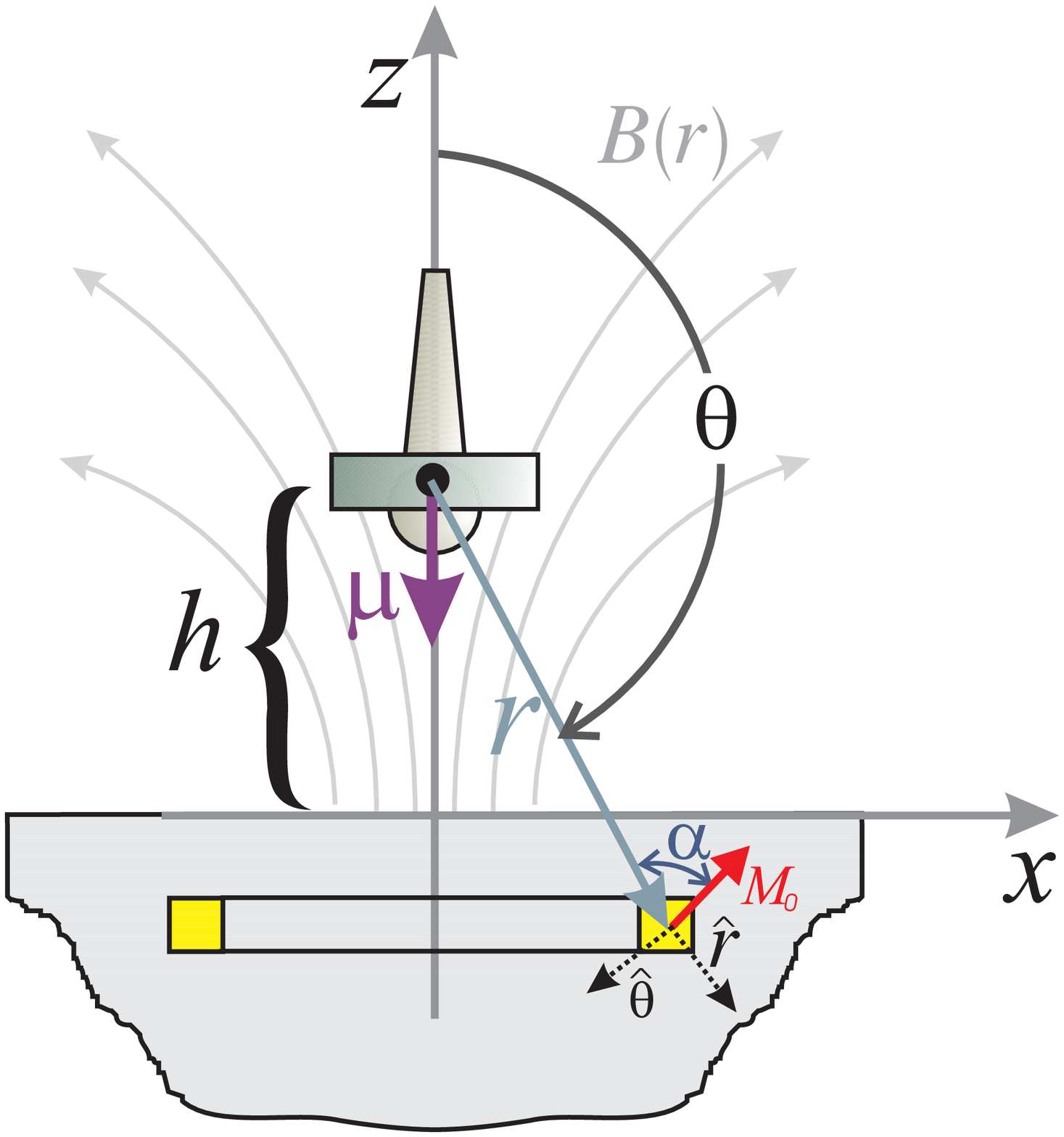}
\end{center}
\caption{The magnetization density at point $(r,\theta)$ is $M_{0}$ and its
direction makes an angle $\alpha\left(  r,\theta\right)  $ with the unit
vector $\mathbf{\hat{r}}$. The angle distribution-$\alpha\left(
r,\theta\right)  $ is chosen so as to maximize the levitation height while
maintaining the translational stability of the top. }%
\label{fig5}%
\end{figure}

\begin{figure}[th]
\begin{center}
\includegraphics[
height=6.2889in,
width=4.8758in
]{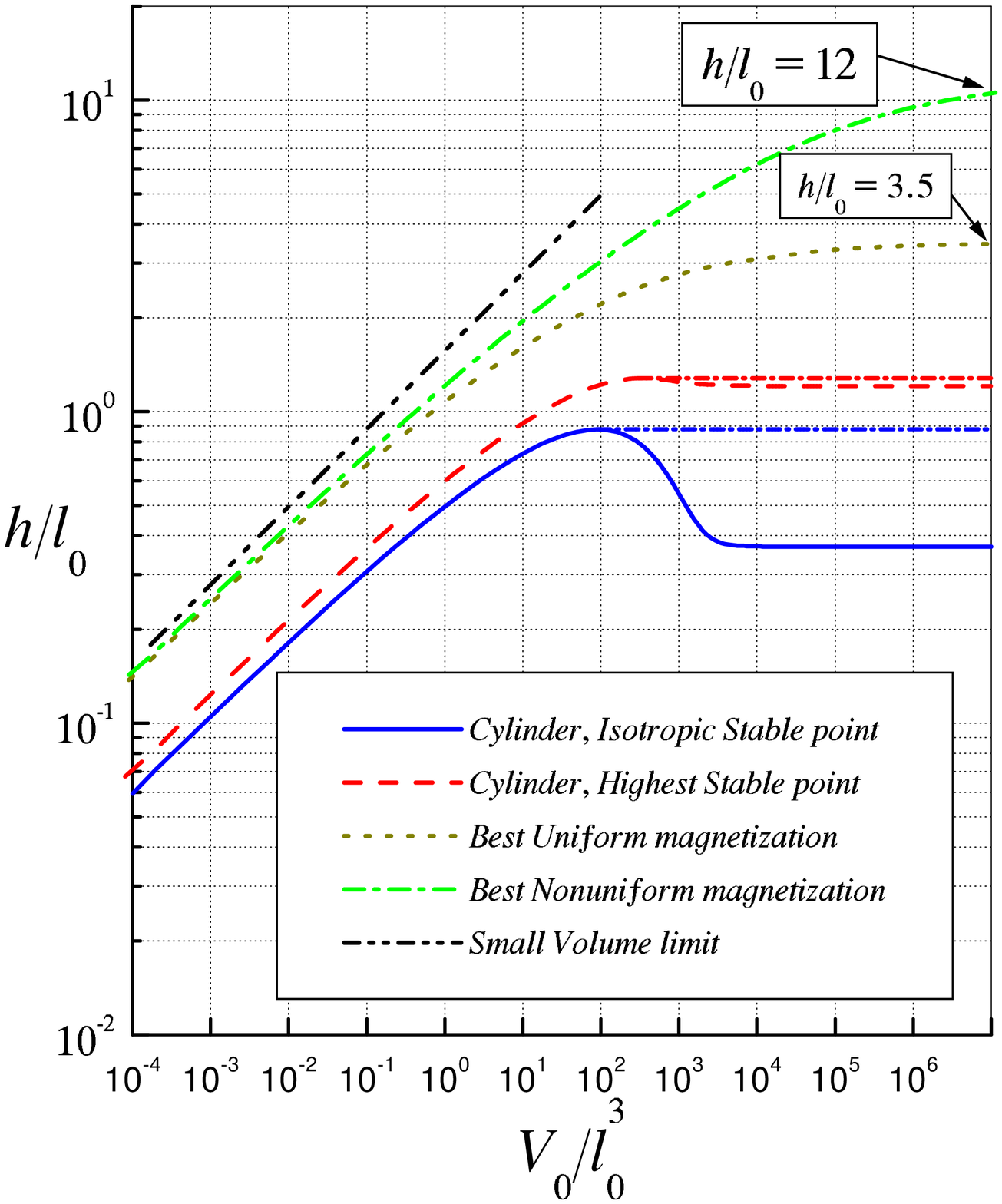}
\end{center}
\par
\caption{Bounds on the maximum levitation height Vs. volume for: Best
magnetization possible (dash-dot line), Best uniform magnetization (dotted
line), Uniformly magnetized cylinder with highest stable point (dashed line),
Uniformly magnetized cylinder with isotropic stable point (solid line) and
small volume limit (dash-dot-dot line). The two curves that correspond to the
uniformly magnetized cylinder has been artificially corrected by extending the
highest levitation point to higher volumes.}%
\label{plots}%
\end{figure}

Consider an infinitesimal magnetic dipole at the point $(r,\theta)$ within the
base with magnetization $M_{0}$. It is situated below the expected hovering
position, as is shown in Fig.(\ref{fig5}), and its direction makes an angle
$\alpha(r,\theta)$ with the line joining the dipole to the equilibrium
position of the top.

Substituting
\[
{\mathbf{M}}=-\sin\alpha\left(  r,\theta\right)  {\mathbf{\hat{\theta}}}%
-\cos\alpha\left(  r,\theta\right)  {\mathbf{\hat{r}},}%
\]
into Eqs.(\ref{eq1.4}),(\ref{eq1.41}), gives
\begin{align}
\left.  \dfrac{\partial B_{z}}{\partial z}\right|  _{z=h}  &  =M_{0}\dfrac
{\mu_{0}}{4\pi}\int\limits_{\pi/2}^{\pi}d\theta\sin\theta%
%TCIMACRO{\dint \limits_{-h/\cos(\theta)}^{\infty}}%
%BeginExpansion
{\displaystyle\int\limits_{-h/\cos(\theta)}^{\infty}}
%EndExpansion
dr\dfrac{2\pi r^{2}}{r^{4}}\label{eq5}\\
&  \times\left\{
\begin{array}
[c]{c}%
-6\sin\alpha\sin\theta\cos\theta\\
-3(3\cos^{2}\theta-1)\cos\alpha
\end{array}
\right\}  ,\nonumber
\end{align}
and
\begin{align}
\left.  \dfrac{\partial^{2}B_{z}}{\partial z^{2}}\right|  _{z=h}  &
=M_{0}\dfrac{\mu_{0}}{4\pi}\int\limits_{\pi/2}^{\pi}d\theta\sin\theta%
%TCIMACRO{\dint \limits_{-h/\cos(\theta)}^{\infty}}%
%BeginExpansion
{\displaystyle\int\limits_{-h/\cos(\theta)}^{\infty}}
%EndExpansion
dr\dfrac{2\pi r^{2}}{r^{5}}\label{eq6.1}\\
&  \times\left\{
\begin{array}
[c]{c}%
-9\sin\theta\left(  5\cos^{2}\theta-1\right)  \sin\alpha\\
-12\cos\theta(5\cos^{2}\theta-3)\cos\alpha
\end{array}
\right\}  .\nonumber
\end{align}

The target function to be extremized in this case, is given by
\begin{equation}
T\left[  \alpha(r,\theta)\right]  =\left.  \frac{\partial B_{z}}{\partial
z}\right|  _{z=h}+\lambda\left.  \dfrac{\partial^{2}B_{z}}{\partial z^{2}%
}\right|  _{z=h}, \label{eq7}%
\end{equation}
and is a functional of $\alpha(r,\theta)$, with $\lambda$ being the Lagrange's
multiplier. Since the variation of $T$ with respect to $\alpha(r,\theta)$ must
vanish, we find, on substitution of Eqs.(\ref{eq5}) and (\ref{eq6.1}) into Eq.
(\ref{eq7}), that $\alpha(r,\theta)$ is given by
\begin{equation}
\tan\alpha(r,\theta)=\frac{2\sin\theta\cos\theta+3\dfrac{\lambda}{r}\sin
\theta\left(  5\cos^{2}\theta-1\right)  }{(3\cos^{2}\theta-1)+4\dfrac{\lambda
}{r}\cos\theta(5\cos^{2}\theta-3)}\text{.} \label{eq7.1}%
\end{equation}

Using Eq.(\ref{eq7.1}) inside Eq.(\ref{eq6.1}), and requiring that
$\partial^{2}B_{z}/\partial z^{2}=0$, gives the equation for $\lambda/h$
\begin{equation}
\int\limits_{\pi/2}^{\pi}d\theta\sin(\theta)%
%TCIMACRO{\dint \limits_{0}^{-\cos(\theta)}}%
%BeginExpansion
{\displaystyle\int\limits_{0}^{-\cos(\theta)}}
%EndExpansion
ydy\left\{
\begin{array}
[c]{c}%
3\sin\theta\left(  5\cos^{2}\theta-1\right)  \sin\alpha\\
+4\cos\theta(5\cos^{2}\theta-3)\cos\alpha
\end{array}
\right\}  =0\text{,} \label{eq9}%
\end{equation}
in which $\alpha(r,\theta)$ depends on $y$ according to
\[
\tan\alpha(r,\theta)=\frac{2\sin\theta\cos\theta+3\dfrac{\lambda}{h}%
y\sin\theta\left(  5\cos^{2}\theta-1\right)  }{(3\cos^{2}\theta-1)+4\dfrac
{\lambda}{h}y\cos\theta(5\cos^{2}\theta-3)}.
\]
Note that in Eq.(\ref{eq9}) the variable $r$ has been changed to $y=h/r$. This
way the double integration is finite and the singularity in the integrand is eliminated.

The numerical solution of Eq.(\ref{eq9}) gives
\[
\lambda=0.424h.
\]
Using this value inside Eq.(\ref{eq5}) yields
\begin{equation}
\left.  \frac{\partial B_{z}}{\partial z}\right|  _{z=h}=-1.33\dfrac{\mu_{0}%
}{4\pi}\frac{2\pi M_{0}}{h}, \label{eq10}%
\end{equation}
which, together with the equilibrium condition, gives
\begin{equation}
h_{\max}\simeq8.4\dfrac{\mu_{0}}{4\pi}\frac{M_{0}\mu}{mg}=8.4l_{0}.
\label{eq11}%
\end{equation}

Note that, though the gradient of the field is finite, the field itself
\emph{diverges}, and hence $k_{\rho}=-k_{z}/2$. As $k_{z}=0$ in the case we
studied here, we conclude that $k_{\rho}$ also vanishes. Thus, the lowest
stable hovering height and the highest one \emph{coincide}, leaving no range
of stability. A small relaxation in the conditions however, such as limiting
the volume of the material to a finite, though very large value, results in a
formation of a finite albeit small, range of stability.

\section{Discussion.\label{sec3}}

We showed that the maximum levitation height of the U-CAS is bounded and is
given in terms of a characteristic length $l_{0}$, which depends on the
properties of the substance that the base and the top are made of. Note also,
that $\mu/m=M_{0}^{top}/\rho$ where $\rho$ is the mass density of the top, and
$M_{0}^{top}$ is its magnetization density. Since $M_{0}\simeq B_{r}/\mu_{0}$,
where $B_{r}$ is the residual induction, and since $B_{r}$ is related to the
energy product \cite{parker} via $(BH)_{\text{max}}\simeq B_{r}^{2}/4\mu_{0}$,
we may also write $l_{0}$ as
\begin{equation}
l_{0}=\dfrac{\mu_{0}}{4\pi}\dfrac{M_{0}^{top}M_{0}^{base}}{\rho g}\simeq
\dfrac{1}{\pi}\dfrac{\sqrt{(BH)_{\text{max}}^{top}(BH)_{\text{max}}^{base}}%
}{\rho g}\text{.} \label{eq1.61}%
\end{equation}
In this form, $l_{0}$ can be easily estimated from the knowledge of the energy
product of the material. The best candidates for large hovering heights are
the Nd-Fe-B magnets. An example of which is Vacuumschmelze's VACODYM 344 HR
\cite{vacuum}, whose remanence ($B_{r}$) is $13.5$ KGauss and whose density is
$7.5$ gr/cm$^{3}$. For this magnet, the magnetization is $M=B_{r}/4\pi=1070$
emu/cm$^{3}$. Thus, with $g=980$ cm/sec$^{2}$, we find that $l_{0}\simeq1.6$
meter. This gives, according to Eq.(\ref{eq11}), a maximum stable hovering
height of about $13$ meters!. Typical values of the energy product, density
and corresponding $l_{0}$, for modern commercial magnets available today are
listed in Table I.\begin{table}[ptb]
\centering%
\begin{tabular}
[c]{|c|c||c|c|c||c|}\hline
\multicolumn{2}{|c||}{Base} & \multicolumn{3}{||c||}{Top} & $l_{0}%
$[m]\\\hline\hline
material & ($BH$)$_{\text{max}}$\cite{strnat2} & material & ($BH$%
)$_{\text{max}}$\cite{strnat2} & $\rho$[Kg/m$^{3}$] & \\\hline\hline
Fe-Nd-B & $320$[KJ/m$^{3}$] & Fe-Nd-B & $320$[KJ/m$^{3}$] & $7500$ &
$1.35$\\\hline
Ferrite & $32$[KJ/m$^{3}$] & Fe-Nd-B & $320$[KJ/m$^{3}$] & $7500$ &
$0.43$\\\hline
Strnat\cite{strnat2} & $800$[KJ/m$^{3}$] & Strnat\cite{strnat2} &
$800$[KJ/m$^{3}$] & $8000$ & $3.2$\\\hline
\end{tabular}
\caption{The value of $l_{0}$ for different base-top material combinations.}%
\label{tab1}%
\end{table}

According to Eq.(\ref{eq1.61}) the characteristic length is proportional to
the \emph{energy product}. Therefore, the larger the energy product is, the
larger $l_{0}$ will be. A question is then asked: what is the highest
conceivable energy product? This question has been already discussed by Strnat
\cite{strnat}, according to which `` ...it seems reasonable to assume that the
best room-temperature energy products will never exceed $\sim100$ MGOe'',
which is about $8\cdot10^{5}$ Joules/m$^{3}$, and is also included in Table I.
These predictions, however, assume that a way might be found to give a fairly
high $H_{ci}$ to any magnetic material.

Table II summarizes the upper and lower bounds for the maximum hovering height
under a selected number of constraints.%

%TCIMACRO{\TeXButton{B}{\begin{table}[tbp] \centering}}%
%BeginExpansion
\begin{table}[tbp] \centering
%EndExpansion%
\begin{tabular}
[c]{|lll|}\hline
\multicolumn{1}{|l|}{Constraints} & \multicolumn{1}{l|}{$%
\begin{tabular}
[c]{c}%
Lower\\
bound
\end{tabular}
$} & $%
\begin{tabular}
[c]{c}%
Upper\\
bound
\end{tabular}
$\\\hline\hline
\multicolumn{1}{|c|}{None} & \multicolumn{1}{c|}{$1.3l_{0}$} &
\multicolumn{1}{c|}{$12l_{0}$}\\
\multicolumn{1}{|c|}{U} & \multicolumn{1}{c|}{$1.3l_{0}$} &
\multicolumn{1}{c|}{$3.5l_{0}$}\\
\multicolumn{1}{|c|}{S} & \multicolumn{1}{c|}{$1.3l_{0}$} &
\multicolumn{1}{c|}{$8.4l_{0}$}\\
\multicolumn{1}{|c|}{U and S} & \multicolumn{1}{c|}{$1.3l_{0}$} &
\multicolumn{1}{c|}{$3.5l_{0}$}\\
\multicolumn{1}{|c|}{V} & \multicolumn{1}{c|}{Fig.(\ref{plots})} &
\multicolumn{1}{c|}{Fig.(\ref{plots})}\\\hline
\end{tabular}
\caption{Summary of the bounds for the maximum levitation height of the U-CAS
under various constraints: U=Uniformly magnetized base plate
, S=stable hovering, V=under a given Volume}%
%TCIMACRO{\TeXButton{E}{\end{table}}}%
%BeginExpansion
\end{table}%
%EndExpansion

It is important to note, that in the above derivation we assumed that the
magnetization vector ${\mathbf{M}}(r,\theta)$ is \emph{independent} of the
field. Though this is a very good approximation for modern permanent magnets
like rare-earth magnets, its only an approximation even for these.
Furthermore, the maximum field at the top is limited by its mechanical and
magnetic strength: Stability considerations show \cite{sync} that the spinning
speed of the top while hovering must be greater than $\sqrt{\mu\left.
B_{z}\right|  _{z=h}/(I_{3}-I_{1})}$ where $I_{3}$ and $I_{1}$ are the moments
of inertia along the principal and secondary axes, respectively. Also, the
maximum field at the top is limited by its coercivity as the field is
\emph{opposite} to the magnetization. Thus, in practice, Eq.(\ref{eq11})
should be considered as an \emph{upper bound} for the maximum hovering height
of the U-CAS.

Yet another way to increase the hovering height of the top is to use current
coil for the base, instead of a permanent magnet \cite{coils}. The advantage
of the coil over the magnetic plate is in the fact that one can raise the top
to the levitation point by electrical means instead of raising the top
mechanically, as in the permanent magnet case. Here, the hovering height is
not limited directly by the strength of the magnetic field, but rather by the
amount of power that one can deliver into the coil to overcome electrical
resistance. One might argue that the use of superconducting wires for the coil
should lift this constraint, but the hovering height is bounded in this case
as well, for if the field increases beyond a critical value, the
superconductor goes into its normal phase. We have shown \cite{coils}, that
for a given height of levitation $h$, the \emph{minimum} power required for
levitation is given by
\[
P=A\frac{rg^{2}h^{3}}{\mu_{m}^{2}}\text{.}%
\]
Here, $P$ is the power in Watts, $r$ is the resistivity in $\Omega\cdot$cm,
$g$ is the free-fall acceleration in cm$\cdot$sec$^{-1}$, $\mu_{m}$ is the
magnetization per unit mass of the top in emu/gr, and $A$ is a number of
dimensions Amp$^{2}$emu$^{2}$/erg$^{2}$cm$^{2}$, and is determined by the
\emph{shape} and \emph{current distribution} of the coil. We found that for a
rectangular cross-section coil, the minimal value of $A\simeq2300$. For the
\emph{optimal} coil however, we find that $A\simeq490$.

\section{Acknowledgments.}

The authors thank Profs. M. Milgrom and M. Kugler for helpful discussions, and
S. Tozik and N. Fernik for help in the calculations. One of the authors (S.
S.) thanks RIKEN, Saitama in Japan and in particular Dr. Y. Kawamura for
kindly hosting him during a short stay in Japan that triggered this study, as
it is then that he became acquainted with the U-CAS.

%\textbf{Shahar Gov} was born in Nes-Ziona, Israel, in 1966. He received the
%B.Sc. degree in Electrical Engineering in 1992 from the Faculty of Electrical
%Engineering of the Technion -- Technological Istitute of Technology, Haife,
%Israel. In 1994 he received his M.Sc. degree in applied physics from the
%Department of Electronics, Weizmann Institute of Science, Rehovot, Israel.
%Currently, he is a Ph.D. student nearing the completion of his thesis work at
%the Weizmann Institute. His current interests include magnetic traps,
%electromagnetic wave propagation and quantum computation.
%
%
%
%
%
%
%
%

\begin{thebibliography}{99}
\bibitem{rem1}Note that $\int_{0}^{1}\sqrt{4\left(  1-x\right)  ^{2}+x^{2}}dx
$ may be computed analytically with the result: $\frac{9}{10}-\frac{2}%
{25}\sqrt{5}\ln\left(  -1+\sqrt{5}\right)  +\frac{2}{25}\sqrt{5}\ln2-\frac
{2}{25}\sqrt{5}\ln\left(  -2+\sqrt{5}\right)  $. In this paper we are only
interested in the numerical result which is $1.24$.
\end{thebibliography}
\end{document}